\documentclass[aps,prc,twocolumn,superscriptaddress,showpacs]{revtex4}
\usepackage{graphicx}% Include figure files
\usepackage{times,mathptmx}
\usepackage{amsmath}
\usepackage{amssymb}

\begin{document}

% Use the \preprint command to place your local institutional report
% number in the upper righthand corner of the title page in preprint mode.
% Multiple \preprint commands are allowed.t
% Use the 'preprintnumbers' class option to override journal defaults
 % to display numbers if necessary
 % \preprint{}

 %Title of paper
% \title{(‰¼) Fusion Hindrance due to Compound-State Formation at 
% Extremely Low Incident Energies in $^{64}$Ni+$^{64}$Ni }
% \title{Existence of One-Body Barrier Revealed in Deep Sub-Barrier Fusion of $^{64}$Ni+$^{64}$Ni }
 \title{Existence of One-Body Barrier Revealed in Deep Sub-Barrier Fusion}
 % repeat the \author .. \affiliation  etc. as needed
 % \email, \thanks, \homepage, \altaffiliation all apply to the current
 % author. Explanatory text should go in the []'s, actual e-mail
 % address or url should go in the {}'s for \email and \homepage.
 % Please use the appropriate macro foreach each type of information

 % \affiliation command applies to all authors since the last
 % \affiliation command. The \affiliation command should follow the
 % other information
% \affiliation can be followed by \email, \homepage, \thanks as well.

\author{Takatoshi Ichikawa}%
%\email{ichikawa.takatoshi@riken.jp}%
\affiliation{RIKEN, Wako, Saitama 351-0198, Japan}
\author{Kouichi Hagino}
\affiliation{Department of Physics, Tohoku University, Sendai 980-8578, Japan}
\author{Akira Iwamoto}
\affiliation{Japan Atomic Energy Agency, Tokai-mura, Naka-gun, Ibaraki
319-1195, Japan}

%\email[1]{}
%\homepage[]{Your web page}
%\thanks{}
%\altaffiliation{}
%Collaboration name if desired (requires use of superscriptaddress
%option in \documentclass). \noaffiliation is required (may also be
%used with the \author command).
%\collaboration can be followed by \email, \homepage, \thanks as well.
%\collaboration{}
%\noaffiliation

\date{\today}

\begin{abstract} 
Based on the adiabatic picture for heavy-ion reactions, in
which the neck formation in the one-body system is taken into
account, we propose a two-step model for fusion cross sections 
at deep subbarrier energies. 
This model consists of the capture process in the two-body
potential pocket, which is followed by the penetration of the 
adiabatic one-body potential to reach a compound 
state after the touching configuration. 
We describe the former process with the 
coupled-channels framework, while the latter 
with the WKB approximation by taking into account the 
coordinate dependent inertia mass. 
The effect of the one-body barrier is 
important at incident energies below the potential energy 
at the touching configuration. 
We show that this model well accounts for the 
steep fall-off phenomenon of fusion cross sections 
at deep subbarrier energies for the $^{64}$Ni+$^{64}$Ni and
 $^{58}$Ni+$^{58}$Ni reactions. 
\end{abstract}

% insert suggested PACS numbers in braces on next line
\pacs{25.60.Pj, 24.10.Eq, 25.70.Jj}
% insert suggested keywords - APS authors don't need to do this
\keywords{}

%\maketitle must follow title, authors, abstract, \pacs, and \keywords
\maketitle

Heavy-ion fusion reactions at low incident energies 
provide a good opportunity to study the quantum tunneling phenomena 
of many-particle systems. Because of a strong cancellation between the 
repulsive Coulomb interaction and an attractive short range nuclear 
interaction between the colliding nuclei, a potential barrier, 
referred to as the Coulomb barrier, is formed, which has to be
overcome in order for fusion to take place. In heavy-ion reactions, 
because of a strong absorption inside the Coulomb barrier, it has been 
usually assumed that the compound nucleus is automatically formed once 
the Coulomb barrier has been overcome. The coupled-channels (CC) 
approach based on this picture has been successful at energies close
to the Coulomb barrier, where the inner turning point of the Coulomb 
barrier is well outside the touching point of the colliding 
nuclei \cite{DHRS98}. 

Recently, 
fusion cross sections have been measured 
for the first time 
at deep subbarrier energies for medium-heavy mass systems, 
such as $^{64}$Ni+$^{64}$Ni, $^{58}$Ni+$^{58}$Ni
and $^{64}$Ni+$^{89}$Y \cite{jiang04,PhysRevLett.89.052701}.
The experimental data indicate that 
fusion cross sections fall off much faster than the exponential 
energy dependence expected from a usual tunneling
picture, as the incident energy decreases. 
Although it has been argued that this hindrance of fusion cross
sections may be explained if one phenomenologically introduces a 
considerably diffuse nuclear potential ~\cite{hagino:054603}, 
the physical origin of the steep fall-off phenomenon has not yet 
been understood (see also Ref. \cite{DHLN06}). 

At energies well below the Coulomb barrier, the inner turning point 
is comparable to, or even smaller than, the touching point. In that
situation, the frozen density approximation, which has often 
been employed in constructing the internucleus potential \cite{SL79},  
breaks down, and one has to treat explicitly the dynamics 
after the touching configuration. 
In this connection, 
Mi\c{s}icu and Esbensen have recently proposed a potential energy 
with a shallow pocket based still on 
the frozen density approximation~\cite{mis06,mis07}. 
That is, 
the outer region of the potential is constructed with the double 
folding procedure \cite{SL79}, while the 
phenomenological repulsive core due to the 
saturation property of nuclear matter is taken into account in the inner 
region~\cite{mis06,mis07}. It was shown that the CC calculation with 
such shallow potential well reproduces the steep fall-off phenomenon 
for the $^{64}$Ni+$^{64}$Ni reaction \cite{mis06,mis07}. 

The approach of 
Mi\c{s}icu and Esbensen is based on 
the sudden picture for nuclear 
reaction, that is, the reaction takes place so rapidly that the 
colliding nuclei overlap with each other 
without changing their density. 
 However, it is not obvious whether the fusion dynamics 
at deep subbarrier energy is 
close to the sudden limit or to the adiabatic limit, where 
the nuclear reaction is 
assumed to take place much more slowly than the 
dynamical density variation of colliding nuclei. 
Since one would not know a priori which 
approach is more reasonable, it is
important to investigate both the possibilities \cite{HW06}. 

In this paper, we investigate
the adiabatic approach in 
explaining the 
steep fall-off phenomenon of fusion cross sections. 
Notice that both the sudden and the adiabatic approaches 
would lead to a similar result to each other in the region 
where the colliding nuclei do not significantly overlap. 
Our model here is to consider 
the fission-like adiabatic potential energy surface with 
the neck configuration after the colliding nuclei touch to each
other. This one-body potential acts like an inner barrier which 
has to be overcome to reach the compound state. 
It is this residual effect which we would like to discuss 
in connection to fusion cross sections at deep subbarrier energies. 

\begin{figure}
\includegraphics[keepaspectratio,width=\linewidth]{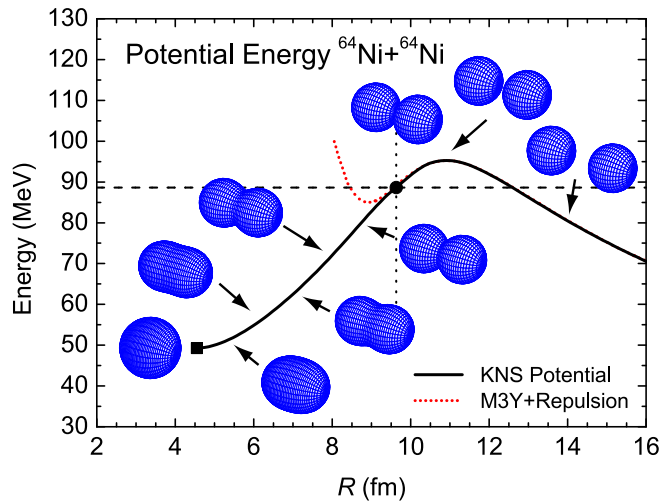}%
\caption{\label{fig1} (Color online) One- and two-body potential energies for
 $^{64}$Ni+$^{64}$Ni obtained with the KNS model as a function of the
center-of-mass distance. The shape for the one-body configuration 
 described by the Lemniscatoids parametrization is also shown. The
 filled circle and square denote the touching 
configuration and the ground state of the compound system,
respectively. 
The dotted lines is the sudden potential 
taken from Ref.~\cite{mis06}. 
}
\end{figure}

In order to illustrate how the adiabatic approach works, 
Fig.~\ref{fig1} shows the potential energy for the 
$^{64}$Ni+$^{64}$Ni reaction obtained with the Krappe-Nix-Sierk (KNS)
model~\cite{kr79} as a function of the center-of-mass distance $R$. 
In the KNS model, the saturation property of nuclear 
matter is phenomenologically taken into account. 
It has also been shown that 
the KNS model is consistent with the potential obtained 
with the energy density formalism 
with the Skyrme SkM$^{*}$ interaction~\cite{de02}. 
The parameters in the KNS model are taken as $a_{0}$=0.68 fm,
$a_{s}$=21.33 MeV and $\kappa_{s}$=2.378 from
FRLDM2002~\cite{moller:072501}. 
The radius parameter is fine-tuned as $r_{0}=1.204$ fm in order to fit
the experimental fusion cross section at high incident energies.
The touching configuration is denoted by the filled circle in the
figure. For distances larger than the touching point, 
the potential energy for the two-body system 
is calculated as the sum of the Coulomb energy for 
two point charges and the nuclear energy given by Eq.~(17) in
Ref.~\cite{kr79}.
For the one-body system after touching two nuclei, 
we assume that the shape
configuration is described by the Lemniscatoids
parametrization (see the inset in the figure)~\cite{lemni82}, and calculate 
the Coulomb and surface integrals for each configuration~\cite{kr79}.

We find that
the value of the potential energy at
the touching configuration $V_{\rm touch}$ is 88.61~MeV.
This is exactly the energy $E_{s}$ at which the experimental
fusion cross section start to fall off abruptly in this
reaction~\cite{jiang04}.
This strongly suggests
a correlation between the
observed fusion hindrance and
a process after the two nuclei overlap each other.
For a comparison, the sudden potential which
Mi\c{s}icu and Esbensen considered \cite{mis06} is denoted by the
dotted line in the figure.
We find that the adiabatic KNS potential and the sudden
potential almost coincide with each other outside the
touching radius.

In order to describe the
two-body process from a large distance to the touching point,
we employ the standard CC formalism by taking into account inelastic
excitations in the colliding nuclei.
However, it is not straightforward to extend this treatment
to the one-body process.
In the CC formalism, the total wave function is expanded with
the asymptotic intrinsic states of the isolated nuclei,
in which one usually restricts the model space only to
those states which are coupled strongly to
the ground state.
Apparently, such asymptotic basis is not efficient to represent the
total wave function for the one-body di-nuclear
system, and in principle one
would require to include all
the intrinsic states in the complete set.
This is almost impossible in practice.
Moreover, the adiabatic one-body potential with the neck
configuration already includes a large part of the channel
coupling effects, and the application of the standard CC formalism
would result in the double counting of the CC effect.

In order to avoid these difficulties,
we here
propose a simple phenomenological model, in which
the two- and one-body processes are defined independently
and time-sequentially.
The fusion cross section in this two-step model then reads
\begin{eqnarray}
\sigma(E)=\frac{\pi \hbar^{2}}{2\mu E} \sum_{\ell}
 (2\ell+1)\,T_{\ell}(E)P_{\rm 1bd}(E,\ell),
\label{two-step}
\end{eqnarray}
where $\mu$ and $E$ denote the reduced mass and the incident
energy in the center-of-mass system, respectively.
$T_{\ell}$ is the capture probability for the two-body process
estimated with the CC method.
$P_{\rm 1bd}$ is the penetrability for the adiabatic
one-body potential to reach the compound state
after the touching of two-body potential, which
plays an important role at energies below
$V_{\rm touch}$ ({\it i.e.,} below the dashed line in Fig.~\ref{fig1}).
At these energies, the fusion reaction is not described
only by the two-body potential, but the potential
which governs the fusion dynamics is switched from the
two-body to the adiabatic one-body potential at the touching
configuration.
Only after overcoming
(or penetrate through) these two- and one-body barriers,
the system can form a compound nucleus.
One may regard the one-body penetrability
$P_{\rm 1bd}$ as a {\it fusion spectroscopic factor}, which
describes the overlap of wave function
between the scattering and the compound states. 

\begin{figure}
\includegraphics[keepaspectratio,width=\linewidth]{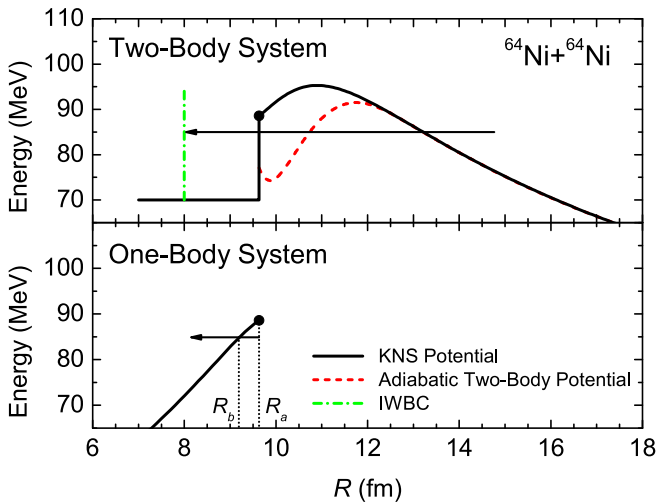}%
\caption{\label{fig2} (Color online) 
The internucleus potential used in the two-step model. The solid line
in the upper panel denotes the KNS potential for the two-body 
process, which is cut at the touching
configuration, while the dashed 
line denotes the lowest two-body eigen potential. 
The dash-dotted line denotes the position at which the 
in-coming wave boundary condition (IWBC) is imposed 
in the CC calculation. 
The solid line in the lower panel denotes the adiabatic one-body
potential inside the touching distance. 
}
\end{figure}

In order to estimate the capture probability $T_{\ell}$ within the
two-step model, 
we cut the two-body potential 
at the touching configuration as shown in the upper panel of
Fig.~\ref{fig2}. 
The capture probability does not depend strongly on how to cut 
the potential, since only the lowest two-body eigen potential, 
which is obtained by diagonalising the coupling Hamiltonian
~\cite{DHRS98,bal98,dasso1983cce}, is relevant at deep subbarrier energies. 
As indicated by the dashed line in the figure, 
the inner turning point for the lowest eigen potential 
is still far outside the touching distance.
Thus, the actual shape of the original potential in the inner-barrier region
influences little on the penetrability. 
Another view is that the incoming wave boundary condition (IWBC) 
is imposed in the CC calculation 
at the touching distance so that 
the capture probability is defined at the touching configuration, 
although in the actual calculations 
we impose the IWBC at a distance somewhat smaller than the touching
point in order to avoid the numerical error. 
For simplicity, we employ a sharp cut-off of the two-body potential in 
this paper. 
\begin{figure}
\includegraphics[keepaspectratio,width=\linewidth]{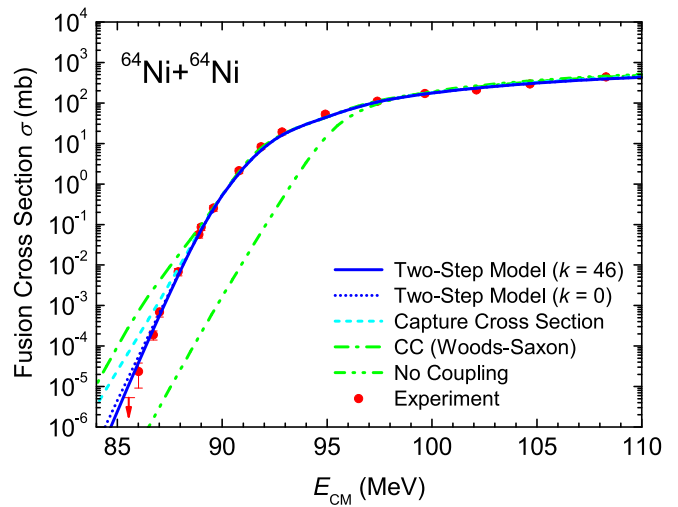}%
\caption{\label{fig3} (Color online) Fusion cross sections for the 
 $^{64}$Ni+$^{64}$Ni reaction calculated with the two-step model.
The filled circles denote the experimental fusion cross section, taken
from Ref. \cite{jiang04}. 
The solid and dotted lines denote the
fusion cross section obtained with the normalization factor 
for the mass inertia of $k=46$ and 0, respectively.
The dashed line denotes the corresponding capture cross section. 
The dash-dotted line is obtained with the Woods-Saxon potential, while 
the dash-dot-doted line shows the result in the absence of the 
channel coupling effect. 
}
\end{figure} 

In order to estimate the one-body probability $P_{\rm 1bd}$, we use the WKB
approximation.
We assume that the reflected flux in this process does not return to
the two-body system, but exits through the muti-dimensional potential
energy surface in the one-body system.
The penetrability then reads 
$P_{\rm 1bd}(E,\ell)=e^{-2S(E,\ell)}$, where
$S(E,\ell)$ is the action integral with the coordinate 
dependent inertia mass $M(R)$, 
\begin{eqnarray}
S(E,\ell)=\int_{R_a}^{R_b}dR\,\sqrt{\frac{2M(R)}{\hbar^2}
(E-V_{\rm 1bd}(R,\ell))}. 
\end{eqnarray}
Here, 
$R_a$ and $R_b$ are the inner and the outer turning points,
respectively (see the lower panel of Fig.~\ref{fig2}). 
$V_{\rm 1bd}$ is the adiabatic one-body potential energy given by  
\begin{eqnarray}
V_{\rm 1bd}(R,\ell)=V_{C}(R)+V_{S}(R)
+\frac{\ell(\ell+1)\hbar^2}{2I(R)}+\frac{2}{7}\,E_{R}, 
\label{pot}
\end{eqnarray}
where $V_{C}$, $V_{S}$ and $I$ are the Coulomb and the surface energies
and the moment of inertia for the rigid body, respectively. 
$E_{R}$ denotes the centrifugal energy at the touching
configuration. Note that the last term in
Eq.~(\ref{pot}) comes from the conservation of the energy and angular
momentum between the two- and one-body systems in the sticking limit~\cite{sticking}.

We now apply the present two-step model to 
the fusion reaction of 
the $^{64}$Ni+$^{64}$Ni system. 
To this end, we use the KNS potential energy
already shown in Fig.~\ref{fig1}. 
In the energy region discussed in this paper, we expect that the
the Lemniscatoids parametrization provides a 
reasonable approximation, because 
the neck formation is still small 
as shown in the inset of Fig.~\ref{fig1}. 
This parametrization has an advantage in that 
the configuration is described with only one parameter
for a symmetric system. 
In addition, 
one obtains a
smooth connection between the one- and two-body potential energy
curves, since the change of the configuration shape 
across the touching point is rather natural. 
As for the inertia mass $M$, 
we take the linear combination between the 
irrotational-flow mass in the Werner-Wheeler
approximation~\cite{PhysRevC.13.2385}, $M_{0}$, and the 
reduced mass, $\mu$. That is, 
$M(R)=k\,(M_{0}(R)-\mu)+\mu$, where $k$ is the normalized factor. 
The renormalization factor is necessary, since 
the liquid drop model with 
the irrotational-flow mass $M_{0}$ overestimates 
the vibrational excitation energy $\hbar\omega_{0}$ for the 
first $2^{+}$ state~\cite{bm75}. 
In the calculations presented below, we use the 
normalization factor, $k=46$, which leads to 
the vibrational energy of 0.2~$\hbar \omega_{0}$.  
Notice that the inertia mass $M$ is in agreement with the reduced mass
$\mu$ at the touching configuration. 

In order to compute the 
capture probability $T_{\ell}$ with the CC framework 
with a sharp-cut KNS potential, 
where the form of the coupling potential is not known, 
we modify the
computer code {\tt CCFULL}~\cite{ccfull} and estimate the nuclear
coupling term with the numerical derivative of the nuclear potential up
to the second order.  
The coupling scheme included in the calculations, as well as the deformation 
parameters, are the same as 
in Ref.~\cite{jiang04}. To be more specific, 
we include the coupling to the low-lying 2$^+$
and 3$^-$ phonon states, 
two-phonon quadrupole excitations, and all possible mutual
excitations both in target and projectile nuclei. 
The potential depth in the inner-barrier region for the sharp-cut 
KNS potential and the position of the
IWBC are chosen as $V_{0}=70$~MeV and $R_{\rm IWBC}=8.0$~fm,
respectively. 
These values are determined using the 
the Woods-Saxon (WS) potential with $V_{\rm WS}$=75.98~MeV, 
$r_{\rm WS}$=1.19~fm, and 
$a_{\rm WS}$=0.676~fm. We have checked 
the numerical stability 
of the calculations at 
extremely low incident energies by comparing the obtained result 
with the one in the 
multi-channel WKB approximation ~\cite{hagi03}. 

Figure 3 shows the fusion cross sections thus obtained. 
It is remarkable that the fusion cross section obtained with $k=46$ for the 
coordinate dependent mass
is in good agreement with the experimental data (see the solid
line). 
The corresponding capture cross sections, obtained by setting 
$P_{\rm 1bd}=1$ in Eq. (\ref{two-step}), is 
denoted by the dashed line. 
As a comparison, the result with the WS potential is also
shown by the dash-dotted line. 
We see that the discrepancy between the capture cross section 
obtained with the WS potential and the
experimental data 
is improved by taking into account the saturation 
property simulated by the KNS potential, and a further improvement 
has been achieved by taking into account the one-body barrier 
inside the touching configuration. 
The result with $k=0$ is denoted by the dotted line. 
The difference between the solid and the dotted line is
small, indicating the negligible effect of the coordinate dependence of
mass inertia in the energy region discussed in this paper.
We have applied the two-step model also to the $^{58}$Ni+$^{58}$Ni
system. We found that the agreement with the experimental
excitation function~\cite{PhysRevC.23.1581} is as good as for the $^{64}$Ni+$^{64}$Ni
system shown in Fig.~\ref{fig3}.

The present two-step model is in the opposite limit to the 
recent sudden model of 
Mi\c{s}icu and Esbensen ~\cite{mis06,mis07}. As long as the fusion 
cross sections are concerned, both the models provide similar 
results, at least for the 
$^{64}$Ni+$^{64}$Ni reaction. However, the origin for the 
fusion hindrance is different between the two approaches. 
In our two-step model, the fusion hindrance takes place due to the 
penetration of the inner one-body potential. On the other hand, 
in the sudden model, which uses a shallow potential, 
the hindrance occurs 
because of the cut-off of the high
angular-momentum components in the fusion cross section. 
The average angular momentum of the
compound nuclei estimated with the sudden model 
would therefore be much smaller than that of the present adiabatic model. 
It is thus interesting to measure 
the average angular
momentum of the compound nucleus at deep subbarrier energies, 
in order to discriminate the two approaches.  

We would next like to 
comment on the recent experimental data for
$^{16}$O+$^{197}$Au, 
where the fusion hindrance was not observed~\cite{back:331}. 
We estimate the potential 
energy at the touching configuration, $V_{\rm touch}$, to be 68.23~MeV
if we use $r_0$=1.2 fm in the KNS potential.  
This is nearly equal to the lowest
incident energy performed in the experiment. Thus, 
the fusion cross sections have to be measured at lower energies in
order to observe the fusion hindrance for this system, as has been 
speculated in Ref. ~\cite{back:331}. 

To summarize, we have proposed the adiabatic two-step model 
for fusion cross sections at deep subbarrier energies. 
By applying this model to the $^{64}$Ni+$^{64}$Ni and $^{58}$Ni+$^{58}$Ni
reactions,
we have shown that the penetration of the 
adiabatic one-body potential with the neck configuration 
after the touching of two colliding 
nuclei is responsible for the 
steep fall-off of fusion cross sections observed 
recently in the experimental data. 
The effect of the one-body potential is important only at energies 
below the potential energy at the touching configuration. 
In this way, the two-step model provides a natural origin for the 
threshold energy of fusion hindrance discussed in
Refs.~\cite{jiang04,PhysRevLett.89.052701}. 

In Ref. \cite{DHLN06}, it was shown that the experimental 
fusion cross sections for the 
$^{16}$O+$^{208}$Pb system follow the exponential energy dependence 
at deep subbarrier energies. This is in contrast to the 
behaviour in the medium-heavy systems discussed in
Refs.~\cite{jiang04,PhysRevLett.89.052701}. 
It would be an interesting future work to apply the present 
two-step model to this  
reaction and to clarify the difference between the 
mass asymmetric and symmetric systems. 

\begin{acknowledgments}
We thank H. Esbensen for discussions on their sudden approach. 
K.H. thanks M. Dasgupta and D.J. Hinde for discussions. 
This work was supported by the Grant-in-Aid for Scientific Research,
Contract No. 16740139 from the Japanese Ministry of Education,
Culture, Sports, Science, and Technology.
\end{acknowledgments}

%\bibliography{hindrance}

\end{document}